\documentclass[reprint,amsmath,amssymb,aps,superscriptaddress,twocolumn,10pt]{revtex4-1}
\usepackage[colorlinks=true,allcolors=blue]{hyperref}
\usepackage{graphicx,color}
\usepackage{dcolumn}
\usepackage{subfigure}
\usepackage{bm}
\usepackage{amsmath,amsfonts,amssymb}
\usepackage{acronym}
\usepackage{enumitem}
\usepackage{array}
\usepackage{multirow}
\allowdisplaybreaks
\newcommand{\beq}{\begin{equation}}
\newcommand{\edeq}{\end{equation}}
\newcommand{\beqn}{\begin{eqnarray}}
\newcommand{\edqn}{\end{eqnarray}}

\newcommand{\SOUTHCUT}{
School of Physics and Optoelectronics, South China University of Technology,\\
 Guangzhou 510641, People's Republic of China}
\newacro{EMRI}{extreme-mass-ratio inspiral}
\newacro{MBH}{massive black hole}
\newacro{MCO}{massive compact object}
\newacro{MECO}{massive exotic compact object}
\newacro{MHO}{massive horizonless object}
\newacro{DWD}{double white dwarf}
\newacro{AK}{analytic kludge}
\newacro{NK}{numerical kludge}
\newacro{AAK}{augmented analytic kludge}
\newacro{CO}{compact object}
\newacro{SNR}{signal-to-noise ratio}
\newacro{PN}{post newtonion}
\newacro{FIM}{Fisher information matrix}
\newacro{LSO}{last stable orbit}
\newacro{GW}{gravitational wave}
\newacro{BH}{black hole}
\newacro{BBH}{Binary Black Hole}
\newacro{BNS}{Binary Neutron Star}
\newacro{NS}{Neutron Star}
\newacro{ISCO}{inner stable circle orbit}
\newacro{SNR}{signal-to-noise ratio}
\newacro{GR}{general relativity}
\newacro{KN}{Kerr-Newmann}
\begin{document}
\title{\textbf{Gravitational waves from extreme-mass-ratio inspirals in the semiclassical gravity spacetime}}
\author{Tieguang Zi}
\email{zitg@scut.edu.cn}
\author{Peng-Cheng Li}
\email{pchli2021@scut.edu.cn, Corresponding author}
\affiliation{\SOUTHCUT}


\begin{abstract}
More recently, Fernandes \cite{Fernandes:2023vux} discovered analytic stationary and axially-symmetric black hole solutions within semiclassical gravity, driven by the trace anomaly. The study unveils some distinctive features of these solutions.
In this paper, we compute the gravitational waves emitted from
the \ac{EMRI} around these quantum-corrected rotating black holes using the kludge approximate method.
Firstly, we derive the orbital energy, angular momentum and fundamental frequencies for orbits on the equatorial plane. We find that, for the gravitational radiation described by quadrupole formulas, the contribution from the trace anomaly only appears at higher-order terms in the energy flux when compared with the standard Kerr case.
Therefore, we can compute the EMRI waveforms from the quantum-corrected rotating black hole
using the Kerr fluxes.
We assess the differences between the EMRI waveforms from rotating black holes with and without the trace anomaly by calculating the dephasing and mismatch. Our results demonstrate that  space-borne gravitational wave detectors can distinguish
the EMRI waveform from the quantum-corrected black holes with a fractional coupling constant of $\sim 10^{-3}$
within one year observation. Finally, we compute the constraint on the coupling constant using the Fisher information matrix method
and find that the potential constraint on the coupling constant by LISA can be within the error $\sim 10^{-4}$ in suitable scenarios.

\end{abstract}
\maketitle
\section{Introduction}
The continuously rising occurrences of gravitational wave (GW) events resulting from the coalescence of compact binaries have been extensively reported \cite{LIGOScientific:2016aoc,LIGOScientific:2018mvr,LIGOScientific:2020ibl,LIGOScientific:2021djp}.
Various issues have been explored using the GW datasets, encompassing tests of General Relativity (GR) and its extensions \cite{LIGOScientific:2021sio,LIGOScientific:2021rnv}, investigations into cosmic histories \cite{LIGOScientific:2021aug},  examinations of dark matter \cite{LIGOScientific:2021ffg,Nagano:2019rbw}, and more
\cite{LIGOScientific:2020ufj,Patricelli:2021rlc}.
With advancements in current and future detector technologies in the realm of detectability, it becomes possible to explore novel astrophysical effects in the vicinity of dark and mysterious sources in the deep universe \cite{Cardoso:2019rvt}. This progress also enables investigations into the nature of classical and quantized horizons \cite{Cardoso:2017cqb,Cardoso:2016rao,Maggio:2019zyv,LISA:2022kgy}.
Particularly, according to the Kerr hypothesis in the context of GR \cite{Robinson:1975bv}, these GW detections may help us to verify whether the massive compact objects are Kerr black holes (BHs) or not \cite{Barack:2006pq,Cardoso:2016ryw,Zi:2021pdp}.
Besides, the presence of quantized signal of horizon is possible to be explored using the GW echoes observed by space-based GW detectors \cite{Cardoso:2017cqb,LISA:2022kgy}.

The no-hair theorem in the framework of GR supports that the BHs can be described by \ac{KN}
solution without any charge \cite{Newman:1965my,Chrusciel:2012jk}. This, however, can be evaded in the
alternatives of GR (see reviews \cite{Chrusciel:2012jk,Berti:2015itd,Barack:2018yly}).
In those extension theories, BHs can carry extra non-trivial scalar, vector or tenser hairs,
which posses more complex multipolar structure compared with the stationary and axisymmetric Kerr BH.
Additionally, after considering the exotic matter, the no-hair theorem can be circumvented, where
the axisymmetry structure of spacetime is broken  \cite{Liebling:2012fv,Raposo:2020yjy,Sanchis-Gual:2021edp}.
The evitable cases of the no-hair theorem can result from the astrophysical environments, physical beyond GR and quantum corrections.
Among the violation schemes, the semi-classical approach to GR, as an interesting and useful means to find BH
solutions, includes the quantum effect of matter field and the classical spacetime geometry \cite{Hawking:1975vcx}.
One popular trace anomaly method essentially is a one-loop quantum correction effect in the invariant classical theory, which breaks the conformal symmetry \cite{Capper:1974ic,Duff:1993wm}.
This could result in stress-energy tensor of quantum fields with non-trivial trace, which is
related only with the curvature of spacetime, making it a feasible way to study the quantum phenomenon in the gravitational fields \cite{Anderson:2007eu,Mottola:2006ew,Mottola:2016mpl,Yang:2022yvq}.
More recently, analytic, stationary and axisymmetric spinning BH solutions have been obtained by considering the presence of the trace anomaly, which own the non-spherically symmetric event-horizon and violate the Kerr bound  \cite{Fernandes:2023vux}.
Unlike the mass of Kerr BH that is independent of the other spatial coordinates,
the quantum-corrected BH (QCBH) solution has a mass function, that is
\begin{equation}\label{ADM:mass}
	\mathcal{M}(r,\theta) = \frac{2M}{1+\sqrt{1-\frac{8\alpha r \xi M}{\Sigma^3} }},
\end{equation}
where $M$ is the ADM mass of the BH, $\Sigma=r^2+a^2\cos^2\theta$ and $\xi=r^2-3a^2\cos^2\theta$
with a spin parameter $a$, and $\alpha$ is the coupling constant of the trace anomaly.
In the context of this new discovered QCBH,
there are several works to study the different effects of the trace anomaly, including the validity of the weak cosmic censorship conjecture \cite{Jiang:2023gpx}, the computation of BH shadows \cite{Zhang:2023bzv} and the study of the non-rotating and rotating cases with a cosmological constant \cite{Gurses:2023ahu}.
In this paper, we plan to study the detectability of the trace anomaly of the QCBH using the observations of the EMRIs, and put constraint on the coupling constant $\alpha$ with the future space-borne GW observatories, such as LISA \cite{LISA:2017pwj}, TianQin\cite{TianQin:2015yph}
and Taiji \cite{Ruan:2018tsw}.

An EMRI is the dynamic process of inspiral, wherein a stellar-mass compact object (CO) with mass $\mu$ orbits around the massive black hole (MBH) with mass $M$ at the core of a galaxy. During this process, the orbits slowly dampen due to the radiated GWs until reaching the last stable orbit (LSO). The low frequency GW signal emitted from such binaries exactly lies in the $m\rm Hz$ band of LISA, which is expected to be one of the key target sources \cite{Berry:2019wgg,Berti:2019xgr}.
In addition, the EMRI systems carry a tiny mass ratios $\epsilon=\mu/M\in[10^{-7},10^{-4}]$,
which implies that the CO accumulates $\mathcal{O}(1/\epsilon)\sim 10^4-10^7$ in the strong field regime
until it is captured by the MBH. After matched filtering the phase in the LISA data stream with the waveform templates \cite{Gair:2004iv,Amaro-Seoane:2007osp}, EMRI detections allow to accurately measure the parameters of sources or test the Kerr nature of the MBH \cite{Barack:2006pq,Babak:2017tow,Fan:2020zhy,Zi:2021pdp}.
From the point of view of astrophysics, it is possible to probe the astrophysical environment
\cite{Kocsis:2011dr,Barausse:2014tra,Cardoso:2022whc,Dai:2023cft,
Figueiredo:2023gas,Zi:2023omh,Rahman:2023sof,Rahman:2022fay,Kumar:2023bdf}.
Besides, from the viewpoint of detecting the nature of horizons, there have been some works to
study the differences of the presence or absence of horizon in the EMRI waveforms \cite{Sago:2021iku,Maggio:2021uge,Zhang:2021ojz}, which also considered several effects, such as tidal heating \cite{Datta:2019euh,Datta:2019epe}, tidal deformability \cite{Pani:2019cyc,DeLuca:2022xlz,Zi:2023pvl} and area quantization \cite{Agullo:2020hxe,Datta:2021row}.
Based on the above mentioned literature, it is feasible to probe the astrophysical effects near the horizon using the EMRI observations. Thus we intend to compute the EMRI waveform from the QCBHs and analyze the distinction of QCBHs from Kerr BHs by computing the dephasing and
mismatch, then obtain the constraint on the deviation parameter $\alpha$ of QCBHs with \ac{FIM} method.

This paper is organized as follows. In Sec. \ref{method},
the calculation recipe of EMRI waveform in the spacetime of the QCBH is presented,
which includes the review of the spacetime background and geodesics in \ref{method:bg},
the evolution method to compute EMRI inspiral trajectories in Sec. \ref{method:inspire},
waveform formula and analysis method of GW data in Sec. \ref{method:waveform}.
We show the results of waveform comparison and the
constraints using the EMRI observations of LISA  in Sec. \ref{result}. Finally, we give a brief summary in Sec. \ref{summary}.
The geometric units $G=c=1$ throughout this paper are utilized.

\section{Method}\label{method}
\subsection{Background and geodesics}\label{method:bg}
The stationary and axisymmetric BH solutions have been obtained by solving the semiclassical Einstein equations with the type-A trace anomaly. 
The solutions are called as QCBHs, whose line element is written as the following form in ingoing Kerr coordinates \cite{Fernandes:2023vux}
\begin{align}
ds^2=&-\left(1-\frac{2r	\mathcal{M}(r,\theta)}{\Sigma}\right)(dv-a \sin^2\theta d\varphi)^2\nonumber\\
&+2(dv-a \sin^2\theta d\varphi)(dr-a\sin^2\theta d\varphi)
\\&+\Sigma(d\theta^2+\sin^2\theta d\varphi^2).\nonumber
\end{align}
Note that this metric is  of the form of the one in the principled-parameterized approach that implements locality and regularity \cite{Eichhorn:2021etc,Eichhorn:2021iwq}. We would like to work in the more familiar Boyer-Lindquist coordinates. However, due to the fact that the spacetime is non-circular, the usual coordinate transformation between the the ingoing Kerr coordinates and the Boyer-Lindquist coordinates in the Kerr case cannot be applied here, since curvature singularities which lie on the
horizon would be introduced. Refs. \cite{Eichhorn:2021iwq,Delaporte:2022acp} found the proper coordinate transformation that respects non-circularity is given by
\begin{align}
t=&v-\int dr \frac{r^2+a^2}{r^2+a^2-2\mathcal{M}(r,\theta)r},\\
\psi=&\varphi-\int dr\frac{a}{r^2+a^2-2\mathcal{M}(r,\theta)r},
\end{align}
where the coordinates $t,r, \theta, \psi$ form the the standard coordinate system in Boyer-Lindquist form if $\mathcal{M}(r,\theta)$ is a constant. Then the metric of the QCBH can be written as
\begin{align}\label{Kerr::semi}
ds^2 =& g_{tt}dt^2 +2g_{t\psi}dtd\psi +2g_{t\theta}dtd\theta+ g_{rr}dr^2 + g_{\theta\theta}d\theta^2 \nonumber\\
&+2g_{\theta\psi}d\theta d\psi+ g_{\psi\psi}d\psi^2\nonumber\\
=&-\left[1-\frac{2r\mathcal{M}}{\Sigma}\right]dt^2+\frac{\Sigma}{\Delta} dr^2-\frac{4ra\mathcal{M} \sin^2\theta }{\Sigma}dtd\psi\nonumber\\
&+\left[\frac{\Sigma}{\sin^2\theta }-\mathcal{M}_2^2-\sin^2\theta\mathcal{M}_1^2(r^2+a^2)+\frac{2r\mathcal{M}}{\Sigma}\tilde{\mathcal{M}}^2\right]d\chi^2\nonumber\\
&-\left[(r^2+a^2)\sin^2\theta+\frac{2\mathcal{M}ra^2\sin^4\theta}{\Sigma}\right]d\psi^2\nonumber\\
&+\left[\frac{4\mathcal{M}r\tilde{\mathcal{M}}}{\Sigma}-2\mathcal{M}_2 \right]dtd\chi\\
&-\left[\frac{4\mathcal{M} r a \tilde{\mathcal{M}}\sin^2\theta}{\Sigma}
-2\sin^2\theta(r^2+a^2)\mathcal{M}_1 \right]d\chi d\psi,\nonumber
\end{align}
with
\beq
\Delta = r^2 -2\mathcal{M} r +a^2,\quad \chi=\cos\theta,
\edeq
\beq
\mathcal{M}_1(r,\chi)=\frac{d}{d\chi}\int dr\frac{a}{r^2+a^2-2\mathcal{M}(r,\theta)r},
\edeq
\beq
\mathcal{M}_2(r,\chi)=\frac{d}{d\chi}\int dr \frac{r^2+a^2}{r^2+a^2-2\mathcal{M}(r,\theta)r},
\edeq
and
\beq
\tilde{\mathcal{M}}(r,\chi)=\mathcal{M}_1(r,\chi)+\mathcal{M}_2(r,\chi).
\edeq
One can see that above metric is far more complicated than the simple Kerr metric. However, one  can find that if we restrict ourselves to the equatorial plane, the above metric would have a simple connection with the Kerr metric by replacing the mass in the Kerr metric with the mass function $(\ref{ADM:mass})$.
When the parameter $\alpha=0$, the solution returns to the classical Kerr BH.

In this paper, we focus on the orbits and waveform of EMRIs, thus the timelike geodesics of
the QCBH would be presented.
Firstly, we introduce some basics about the geodesics.
For a point particle, the Hamiltonian can be written as
\begin{equation}\label{eq:Hamilton}
\mathcal{H} = \frac{1}{2}g_{\mu\nu}p^\mu p^\nu=-\frac{1}{2}m^2
\end{equation}
where $p^\mu$ and $m$ is the momentum and rest mass of the particle, respectively.
Since the spacetime \eqref{Kerr::semi} are stationary and axisymmetric, the particle along the geodesic has two conserved quantities, i.e., the energy and $z$-component of the angular momentum,
\begin{equation}\label{eq:ELz:g}
\frac{E}{m} =-(g_{tt} \dot{t}+ g_{t\psi} \dot{\psi}) , ~~~~~ \frac{L_z}{m} = g_{t\psi}\dot{t} + g_{\psi\psi}\dot{\psi},
\end{equation}
where the dot denotes differentiation with respect to proper time $\tau$.
We can get two first-order decoupled differential equations about the $t$ and $\psi$ momenta
in light of equations \eqref{eq:ELz:g}, the following form is written as
\begin{equation}\label{eq:tdot:psidot}
\dot{t}= \frac{Eg_{\psi\psi}+ L_z g_{t\psi}}{m(g_{t\psi}^2-g_{tt}g_{\psi\psi})},~~~
\dot{\psi}= \frac{Eg_{t\psi}+ L_z g_{tt}}{m(g_{\psi\psi} g_{tt} -g^2_{t\psi})}.
\end{equation}
In this paper we focus on the equatorial orbits, the coordinate $\theta$ becomes to $\theta=\pi$ and $\dot{\theta}=0$.
According to the Eq. \eqref{eq:Hamilton}, we can obtain the equations of the $r$ and simply the Eqs. \eqref{eq:tdot:psidot}
\begin{eqnarray}
r^2\frac{dr}{d\tau} &=& \pm \sqrt{V_r} = \pm\sqrt{T^2-\Delta[m^2r^2+(L_z-aE)^2]},\\
r^2\frac{d\phi}{d\tau} &=& V_\phi=\frac{aT}{\Delta}+L_z - aE,\\
r^2\frac{dt}{d\tau} &=& V_t = \frac{(r^2+a^2)T}{\Delta} + L_z - aE,\\
\theta(\tau) &=& \pi/2.
\end{eqnarray}
where $T=E(r^2+a^2)-aL_z$. Since we will discuss the bound orbits, whose energy satisfies $0<E<1$.
The bound orbits are described by the energy $E$,  angular momentum $L_z$, semi-latus rectum $p$ and
eccentricity $e$.
For the equatorial eccentric orbit, there exists two turning points, that are the periastron $r_p$ and
the apastron $r_a$,
\begin{equation}
r_p=\frac{p}{1+e},~~~r_a=\frac{p}{1-e}.
\end{equation}
The radial potential $V_r$ satisfies $V_r(r_a)=0$ and $V_r(r_p)=0$,
with this conditions, we can get the expressions $E$ and $x$ by defining $x=L_z-aE$.
Since the function $\mathcal{M}$ dependents on $r$, the full expressions of $E$ and $x$
are lengthy to be to be listed even in the appendix, so
we just put them in the ancillary {\em Mathematica} file.
Using the parameters $p$ and $e$, the radial coordinate can be written as
\begin{equation}\label{param:rtochi}
r(\chi)=\frac{p}{1+e\cos\chi},
\end{equation}
where $\chi$ is a monotonic parameter that varies from 0 to $2\pi$,
it equals to 0 at $r=r_p$ and the parameter $\chi=\pi$ at $r=r_a$.
According to the parameterization method \eqref{param:rtochi}, the geodesic equations can be transformed
as
\begin{eqnarray}
\frac{dr}{d\tau} &=&\frac{\sqrt{ \mathcal{V}_{r\chi}}}{p^{3/2} \mathcal{P}_\alpha}  \,,\label{drdtau}\\
\frac{dt}{d\tau} &=& \frac{ ax (1+e\cos\chi)^2}{p^2} \nonumber\\  &&
  + \frac{(1+e\cos\chi)^2 (p^2+a^2(1+e\cos\chi)^2)}
  {a^2p^2(1+e\cos\chi)^2+p^3(p-4M(1+e\cos\chi)/\mathcal{P}_\alpha)}\,\label{dtdtau}\nonumber\\ &&
  \,\,\times \left(\frac{Ep^2}{(1+e\cos\chi)^2 }-ax \right)\\
\frac{d\phi}{d\tau} &=&(1+e\cos\chi)^2 (aEp \mathcal{P}^2_\alpha+(p-4M+p(\mathcal{P}^2-1)x
\nonumber \\&-&4exM\cos\chi))/(p (a^2 e^2 \mathcal{P}_\alpha^2 \cos ^2(\chi )
\nonumber \\ &+&2 e \cos\chi (a^2  \mathcal{P}_\alpha^2-2 M p)+a^2 \mathcal{P}_\alpha^2
\nonumber \\&+&p (-4 M+p   (\mathcal{P}_\alpha^2-1)+p))) \,,\label{dphidtau}
\end{eqnarray}
where
\begin{eqnarray}
\mathcal{V}_{r\chi} & = &
-2 a E p \left(\mathcal{P}_\alpha^2-1\right) x (e \cos (\chi )+1)^2
\nonumber\\ & -&2 a E p x (e \cos (\chi )+1)^2 -p \left(\mathcal{P}_\alpha^2-1\right) (a e \cos
   (\chi )+a)^2
\nonumber\\ & -&  p (a e \cos (\chi )+a)^2+4 M p^2 (e \cos (\chi )+1)
\nonumber\\ & +&4 M x^2 (e  \cos (\chi )+1)^3
 - p \left(\mathcal{P}_\alpha^2-1\right) (e x \cos (\chi )+x)^2
\nonumber \\ & -& p   (e x \cos (\chi )+x)^2
+  E^2 p^3  \left(\mathcal{P}_\alpha^2-1\right) + E^2 p^3
\nonumber \\ & -& p^3   \left(\mathcal{P}_\alpha^2-1\right)-p^3 \,
\end{eqnarray}
with $\mathcal{P}_\alpha =\sqrt{1+\sqrt{1-\frac{8 \alpha  M (e \cos (\chi )+1)^3}{p^3}}}$.
In term of Eqs.\eqref{dphidtau}, \eqref{drdtau}, \eqref{dtdtau} and
using the relation $dr/d\chi=\frac{e p \sin (\chi )}{(e \cos (\chi )+1)^2}$, we can obtain
\begin{eqnarray}
\frac{dt}{d\chi} &=&\frac{dt}{d\tau}\left(\frac{dr}{d\tau}\right)^{-1} \frac{dr}{d\chi}\,,\\
\frac{d\phi}{d\chi} &=&\frac{d\phi}{d\tau}\left(\frac{dr}{d\tau}\right)^{-1} \frac{dr}{d\chi}\,,
\end{eqnarray}
then the radial period is given by $T_r=\int^{2\pi}_0\frac{dt}{d\chi} $, similarly the azimuthal period is $\Delta\phi=\int^{2\pi}_0\frac{d\phi}{d\chi} $.
In term of the two orbital periods, the radial frequency $\Omega_r$ and azimuthal frequency $\Omega_\phi$ can be written as
\begin{eqnarray}\label{orbital:freq}
\Omega_r =\frac{2\pi}{T_r}, ~~~~~~
\Omega_\phi = \frac{\Delta\phi}{T_r}\,.
\end{eqnarray}
Thus the total orbital frequency $\omega$ is equal to linear combination of radial frequency $\Omega_r$ and azimuthal frequency $\Omega_\phi$, which is related with the orbital phase $\Phi$ by
\begin{equation}
	\omega(t) = \frac{d\Phi}{dt}.
\end{equation}
\subsection{Adiabatic evolution}\label{method:inspire}
The dynamic of EMRI orbits are dominated by the dissipation of energy and angular momentum due to gravitational wave emission.
Under the conditions of adiabatic approximation, the secondary object orbits along the a sequence of geodesics up to the \ac{LSO}.
The change rate of the orbital energy and angular momentum $(\dot{E},\dot{L_z})$ can be given by the balance law
\begin{eqnarray}
\dot{E} = \dot{E}^{\rm GW}, ~~~~~~
\dot{L_z} = \dot{L_z}^{\rm GW}\,,
\end{eqnarray}
where the $(\dot{E},\dot{L_z})^{\rm GW}$ are GW fluxes resulted from the loss of EMRI orbits.
The fluxes $(\dot{E},\dot{L_z})^{\rm GW}$  modified by the parameter $\alpha$ of QCBH can be neglected,
it is because that the contribution of the parameter $\alpha$ is higher order at $\sim\mathcal{O}(1/p)^{4}$ for the weak field approximation.
To illustrate this  point, we argue the energy fluxes differences between the Kerr and QCBH cases
in Append \ref{appA:flux}.
Therefore, the fluxes $(\dot{E},\dot{L_z})^{\rm GW}$ are approximately described by fluxes of Kerr BH \cite{Gair:2005ih}, the detailed expression is placed in Append \ref{appA:flux}.

The evolutions of orbital semilatus rectum and eccentricity are given by the following equations \cite{Glampedakis:2002ya}
\begin{eqnarray}
\dot{p} &=& H^{-1} (-\frac{dE}{de}\dot{L_z} + \frac{dL_z}{de} \dot{E}), \label{evolve:pdot} \\
\dot{e} &=& H^{-1} (\frac{dE}{dp}\dot{L_z} - \frac{dL_z}{dp} \dot{E})  \,, \label{evolve:edot}
\end{eqnarray}
where the dot denotes the time derivative and $H^{-1} = \frac{dE}{dp}\frac{dL_z}{de}- \frac{dE}{de}\frac{dL_z}{dp}$.

To quantitatively assess the effect of the additional parameter $\alpha$ on the EMRI observations by LISA,
we evolve the orbital frequencies with and without $\alpha$, following the their inspiral trajectories during the observational time.
For a given time $t$, the dephasing can be defined by  the integral of the difference of orbital frequencies
\begin{eqnarray}\label{eq:dephsing}
\delta\Psi_{r,\phi}(t) = \int_0^t 2(\Omega_{r,\phi}^{\alpha \neq0}-\Omega_{r,\phi}^{\alpha=0})dt,
\end{eqnarray}
where $\Omega_{r,\phi}^{\alpha}$ are the orbital frequencies in Eq. \eqref{orbital:freq}.
\subsection{Waveform}\label{method:waveform}
With the evolution Eqs. \eqref{evolve:pdot} and \eqref{evolve:edot} at hand,
we can simulate the inspiral trajectories with and without the effect of the trace anomaly using numerical method.
The EMRI orbits obtained from Eqs. \eqref{evolve:pdot} and \eqref{evolve:edot} allow to
calculate EMRI signal \cite{Barack:2003fp,Canizares:2012is},
and the spacetime perturbation far from the source is written as in the transverse-traceless (TT) gauge
\begin{eqnarray}\label{eq:hTT}
h^{\rm TT}_{ij} = \frac{2}{D}\left(P_{il}P_{jm}-\frac{1}{2}P_{ij}P_{lm}\right)\ddot{I}_{lm},
\end{eqnarray}
where $D$ is the luminosity distance from source to detector, $P_{ij}=\delta_{ij}-n_in_j$ is the projection operator,
$n_i$ is the unit vector directing from detector to source, and $\delta_{ij}$ is the Kronecker delta.
The quantity $\ddot{I}_{lm}$ is the second time derivative of mass quadrupole moment,
which is given by  $I^{ij}=\mu r^i(t)r^j(t)$.
The GW polarization modes can be simplified in term of Eq. \eqref{eq:hTT} as
\begin{eqnarray}\label{eq:hphc}
h_+(t) &=& \mathcal{A} \cos\left(2\Phi(t)+2\xi \right)(1+\cos^2\iota),\\
h_\times(t) &=&  -2\mathcal{A} \sin\left(2\Phi(t)+2\xi \right)\cos^2\iota,
\end{eqnarray}
where $\mathcal{A} = 2\mu(M \omega(t))^{3/2}/D$,  $\iota$ is the inclination angle between the orbital angular moment and line of sight.
and $\xi$ is the latitudinal angle.
Under the low-frequency approximate condition, the GW strain measured by detector can be given by
\begin{eqnarray}\label{eq:hphc}
h(t) = \frac{\sqrt{3}}{2}[h_+(t)F_+(t) + h_\times(t)F_\times(t)]
\end{eqnarray}
where the interferometer pattern function $F_{+,\times}(t)$ can be determined by four angels,
which describe the source orientation, $(\theta_s,\phi_s)$,
and the direction of \ac{MBH} spin, $(\theta_k,\phi_k)$ in the ecliptic coordinate \cite{Apostolatos:1994mx,Cutler:1997ta}.

The dephasing provide a preliminary criterion of the  QCBH,
a more accurate assessment is to compute the mismatch between two waveforms from EMRI with and without the additional parameter $\alpha$. The mismatch is defined by
\begin{eqnarray}\label{eq:hphc}
\mathcal{M} = 1 - \mathcal{O}(h_a|h_b),
\end{eqnarray}
where the overlap $\mathcal{O}(h_a|h_b)$ is given by inner product
\begin{equation}
\mathcal{O}(h_a|h_b) = \frac{<h_a|h_b>}{\sqrt{<h_a|h_a><h_b|h_b>}},
\end{equation}
with the noise-weighted inner product $<h_a|h_b>$ is defined by
\begin{align}\label{inner}
<h_a |h_b > =2\int^\infty_0 df \frac{h_a^*(f)h_b(f)+h_a(f)h_b^*(f)}{S_n(f)},
\end{align}
here the tilde and star stand for the Fourier transform and complex conjugation, respectively.
The noise  power spectral density $S_n(f)$ of space-borne GW detector, such as LISA \cite{LISA:2017pwj}.
When two waveforms keep equivalent, the overlap is one and the mismatch is vanishing.
A rule of thumb formula regarding waveforms resolution has been proposed,
in which the detector would distinguish two waveforms if their mismatch satisfies $\mathcal{M}\geq\mathcal{D}/(2\rho^2)$
\cite{Flanagan:1997kp,Lindblom:2008cm},
where $\mathcal{D}$ is the number of the intrinsic parameters for the EMRI system and $\rho$ is \ac{SNR} of EMRI signal.
For the EMRI with  QCBH, the number of intrinsic parameters is 7.
Assuming that the minimum SNR is 20 detected by LISA \cite{Babak:2017tow},
so the threshold value of mismatch distinguished by LISA should be $\mathcal{M}\simeq0.01$.

In order to evaluate  capability of  measuring source parameters with LISA,
we conduct the parameter estimation for EMRI source using FIM method \cite{Vallisneri:2007ev}.
The FIM can be defined by
\begin{equation}
    \Gamma_{ab}=\Big( \frac{\partial h}{\partial \lambda_a} \Big| \frac{\partial h}{\partial \lambda_b} \Big),
\end{equation}
where $\lambda_a$, $a=1,2,...,$ are the parameters appearing in the waveform and  the inner product $(|)$ is  defined by Eq. \eqref{inner}.
For the EMRI signal with higher \ac{SNR},  the variance-covariance matrix can be approximately written as
\beq
\Sigma_{ab}\equiv <\Delta \lambda_a \Delta\lambda_b>=(\Gamma^{-1})_{ab}.
\edeq
The uncertainty of the $a$th parameter $\lambda$ is obtained as
\beq
\delta\lambda_a=\Sigma_{aa}^{1/2}.
\edeq
It is remarkable that the FIM  in the linear signal approximation is applicable \cite{Zi:2022hcc}.
The numerical stability of the inverse FIM is discussed in Appendix \ref{appB:Stability}.
\section{Results}\label{result}
We plan to show  the comparison of the EMRI waveforms from the quantum-corrected Kerr and standard Kerr BHs, and quantify the difference by computing the dephasing and mismatch among these waveforms. Furthermore, we will assess the detectability of such modified EMRI signal with the LISA observations in terms of FIM method.

\begin{figure*}[ht]
\centering
\includegraphics[width=0.95\textwidth]{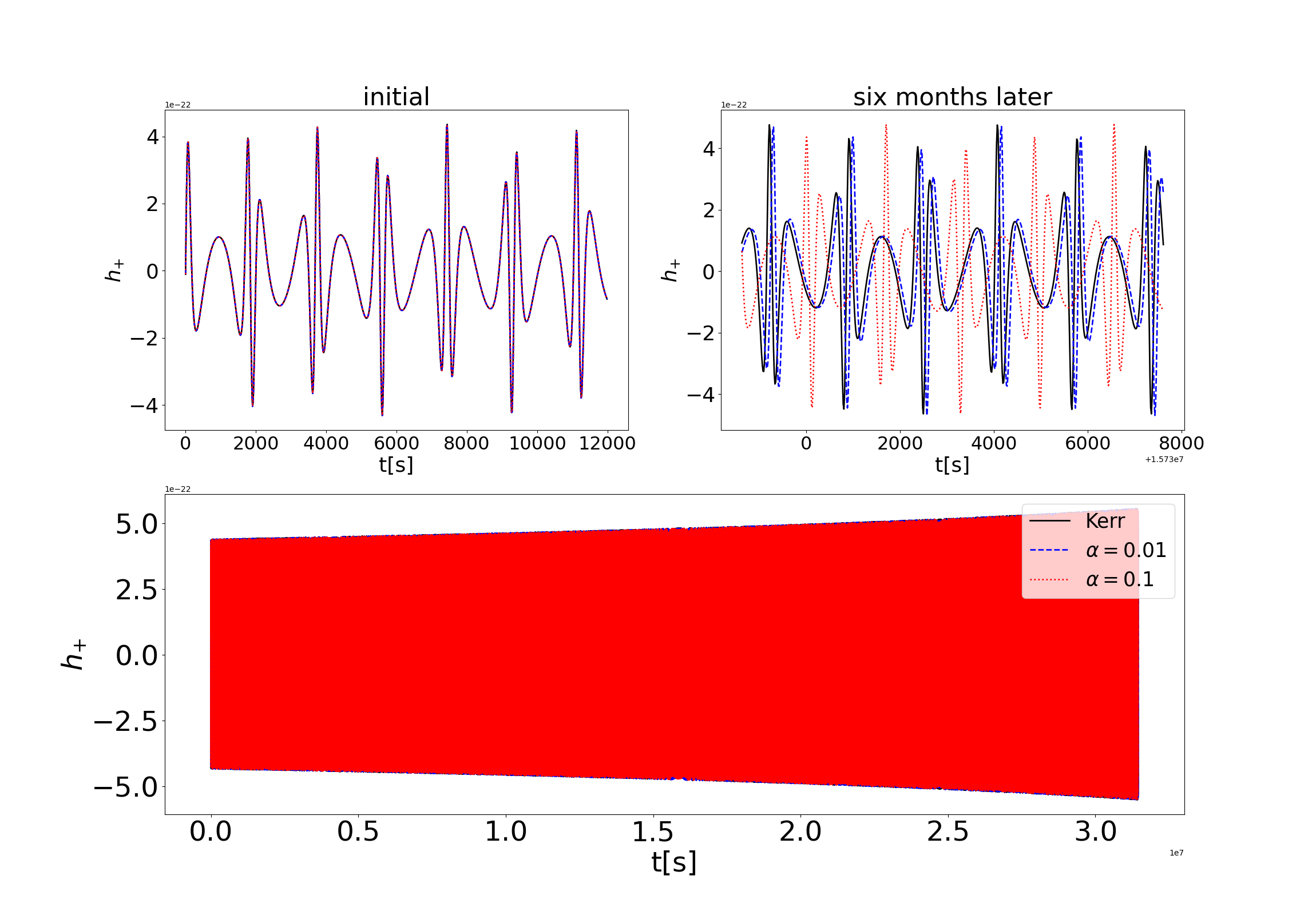}
\caption{Comparison between the polarizations $h_+$ of three EMRI waveforms from the standard Kerr and the quantum-corrected Kerr BHs with derivation parameter $\alpha=(0.01,0.1)$. The BH spin  and the initial orbital parameters
are chosen to $a=0.9$, $p_0=12$ and $e_0=0.5$.
The left panel of  top figures  is  the initial stage of the time domain waveforms
and the right panel of top figures denotes the  time domain waveforms after six months.
The bottom of figure is the full time domain waveforms of $h_+$, where the inspiral time of the \ac{CO} is set as one year.}
\label{fig:waveform}
\end{figure*}

\begin{figure*}[ht]
\centering
\includegraphics[width=0.49\textwidth]{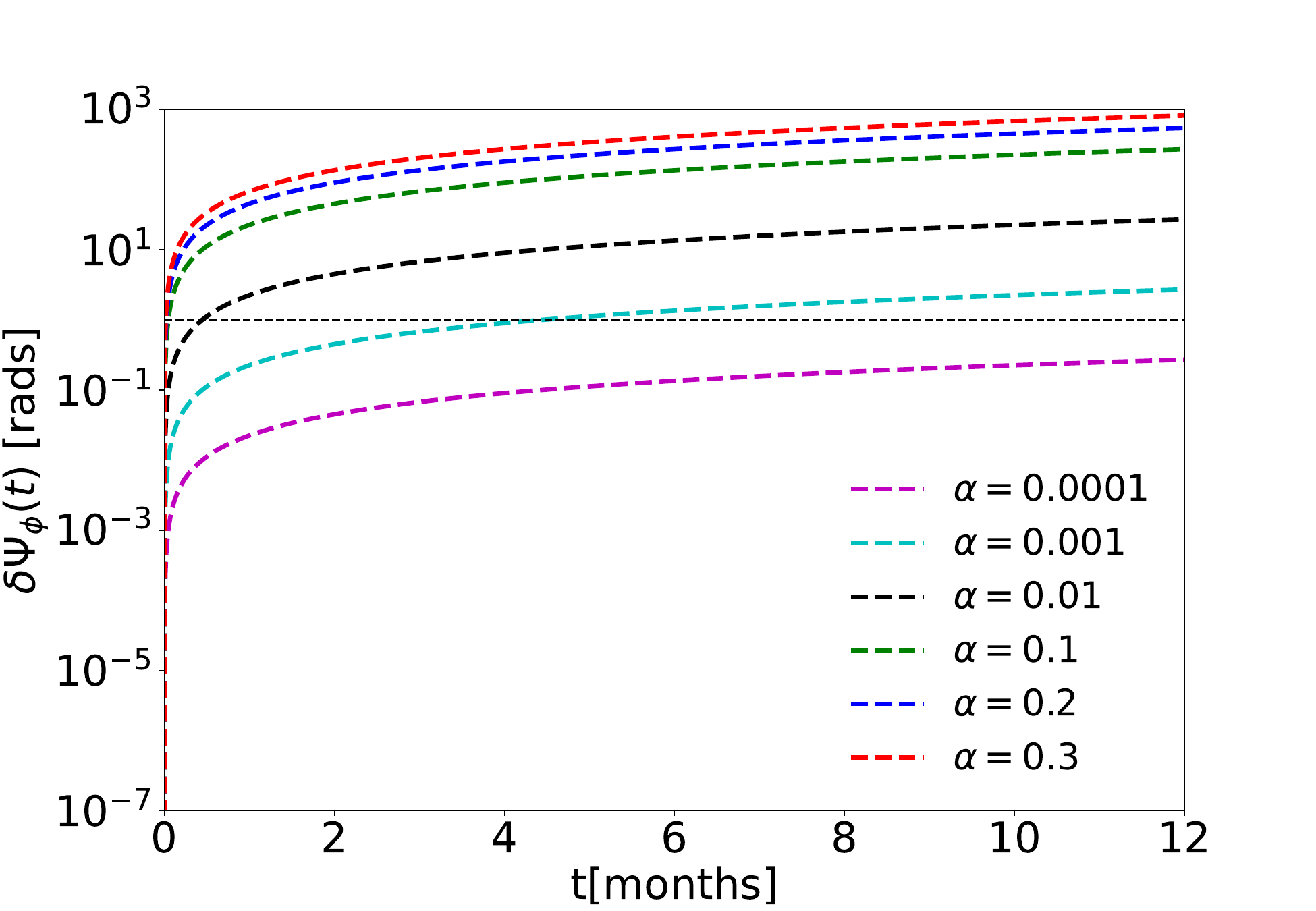}
\includegraphics[width=0.49\textwidth]{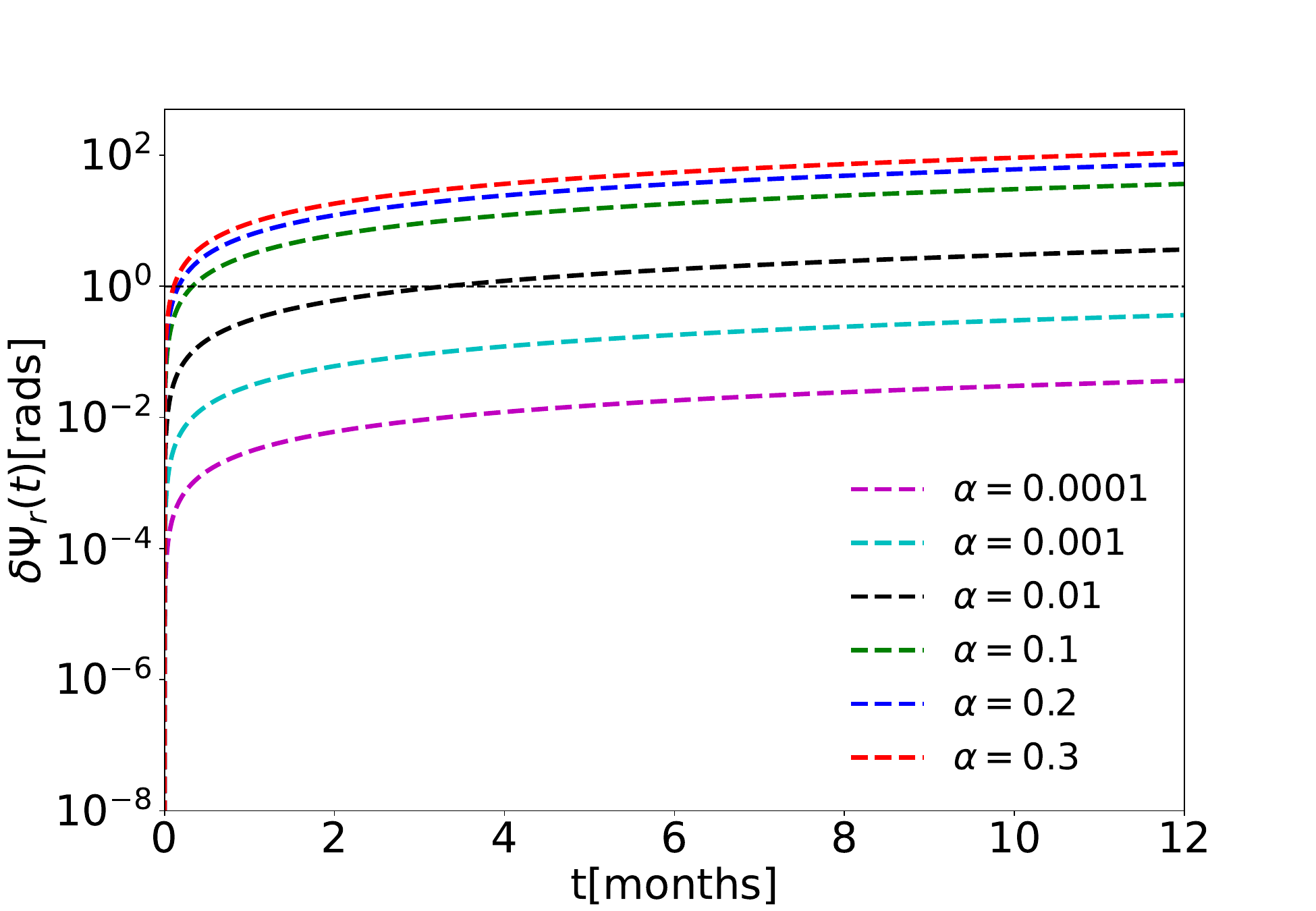}
\caption{Azimuthal (left) and radial (right) dephasings as functions of the observation time, where the spin of MBH  $a=0.9$ and the deviation parameter $\alpha=(0.0001,0.001,0.01,0.1,0.2,0.3)$.
The initial orbital semi-latus rectum and and eccentricity are set to $p_0=12$ and $e_0=0.5$.
The horizontal black dashed line in the figures denotes the threshold for phase that can be distinguished by LISA.
}
\label{fig:Dephasing:alpha}
\end{figure*}

\begin{figure*}[ht]
\centering
\includegraphics[width=0.49\textwidth]{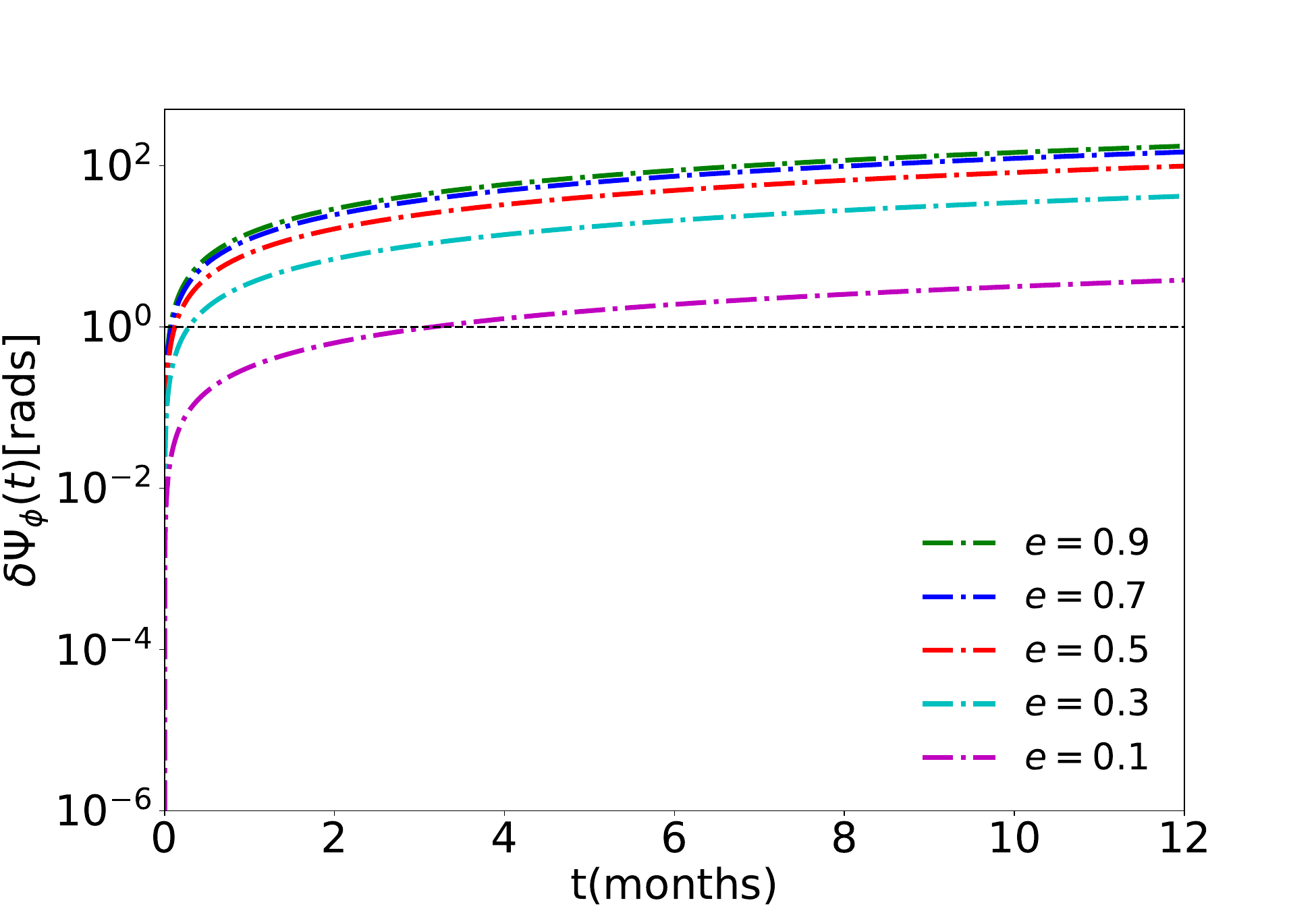}
\includegraphics[width=0.49\textwidth]{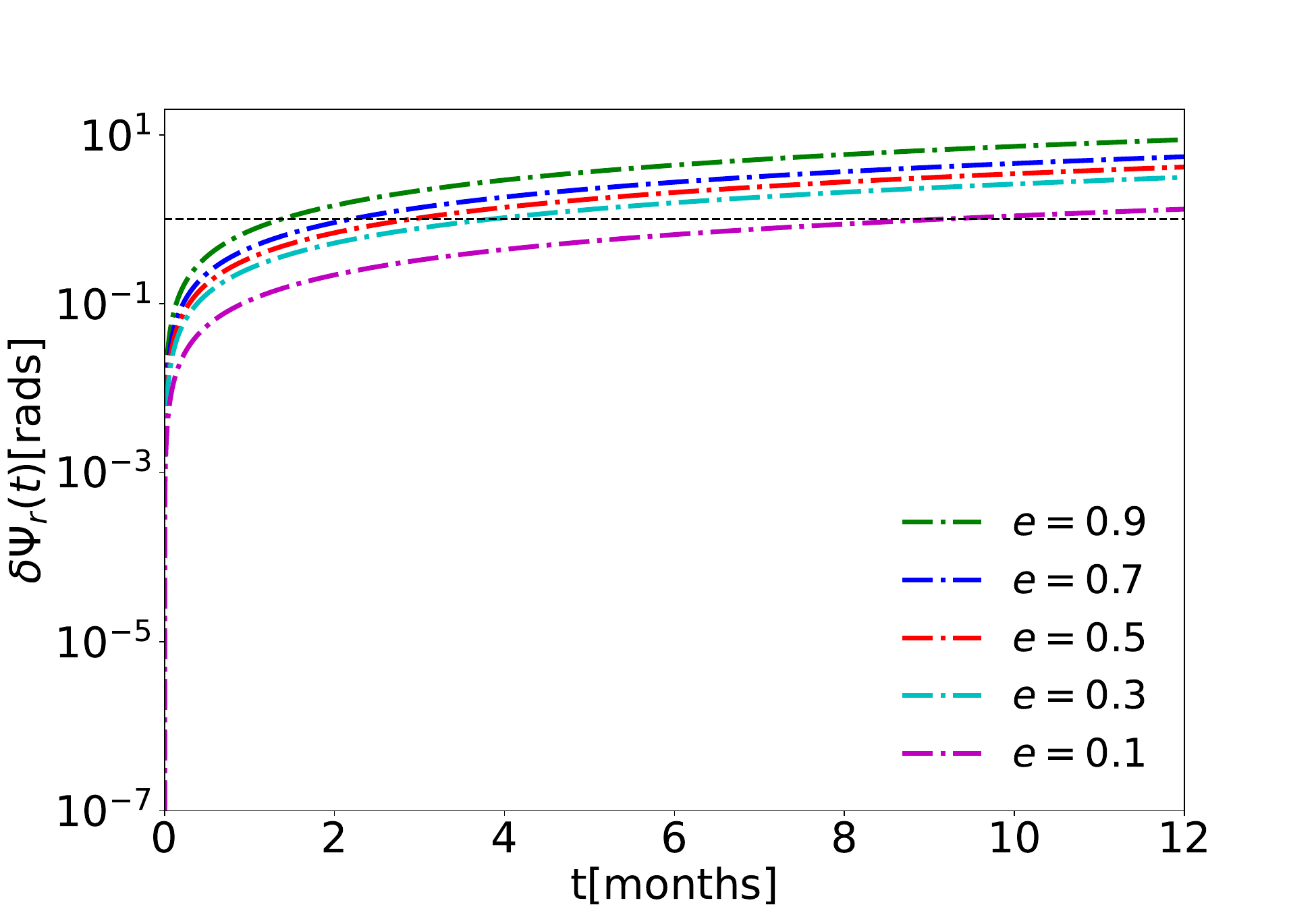}
\caption{Azimuthal (left) and radial (right) dephasings as functions of the observation time are plotted, where the spin of MBH  $a=0.9$ and the deviation parameter  $\alpha=0.01$.
The initial orbital semi-latus rectum and eccentricity are $p_0=12$  and $e_0=(0.1,0.3,0.5,0.7,0.9)$. As before, the horizontal black dashed line in the figures denotes the threshold for phase that can be distinguished by LISA.}
\label{fig:Dephasing:ecc}
\end{figure*}

\begin{figure*}[ht]
\centering
\includegraphics[width=0.49\textwidth]{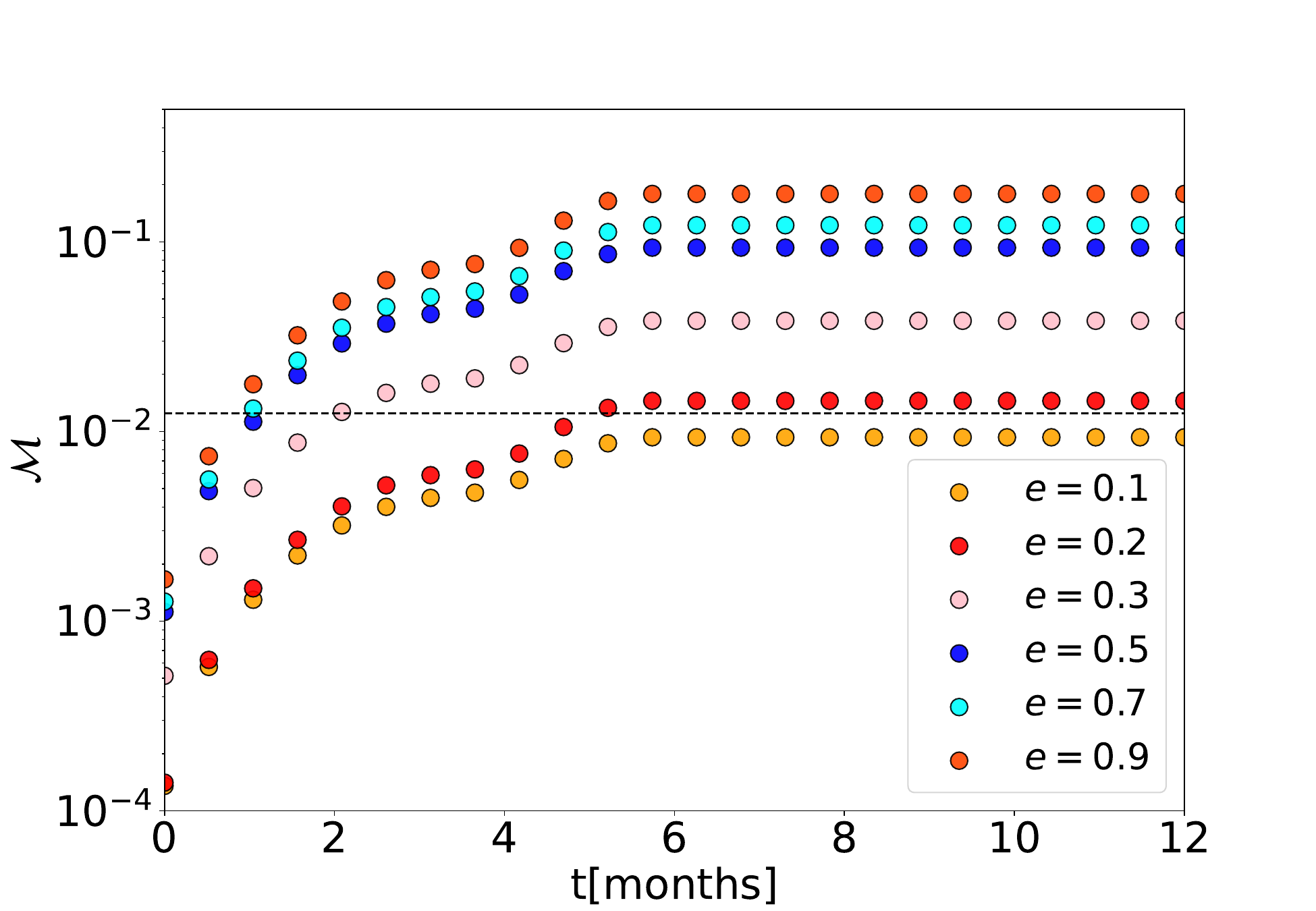}
\includegraphics[width=0.49\textwidth]{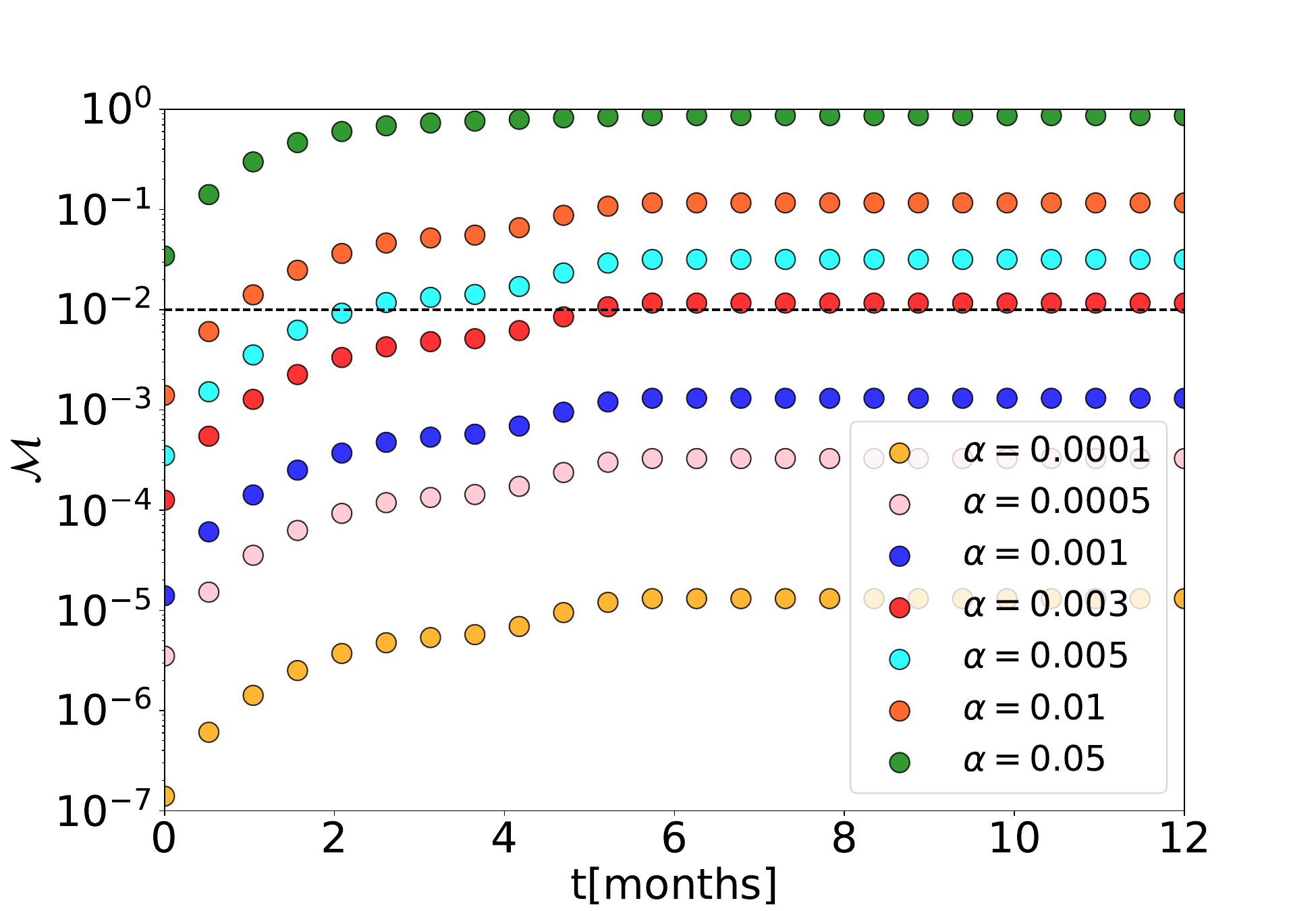}
\caption{Mismatch between waveforms emitted from EMRI systems with and without the trace anomaly  as a function of observation time are plotted,  where the length of waveform is set as one year and $a=0.9$. The left panel depicts several cases of eccentricities $e_0=(0.1,0.3,0.5,0.7,0.9)$ with $\alpha=0.01$ and $p_0=12$, and the right panel shows the examples of different deviation parameters $\alpha=(0.0001,0.0005,0.001,0.003,0.005,0.01,0.05)$ with $e_0=0.5$ and $p_0=12$.
The horizontal black dashed line denotes the minimum value distinguished by LISA.}
\label{fig:mismatch}
\end{figure*}

\subsection{Waveform and mismatch}\label{Waveform}
Firstly, we show the comparison of time domain EMRI waveforms  in the cases with and without the quantum correction in Fig. \ref{fig:waveform}.
Here, we consider the obseravation time for a \ac{CO} inspiraling into the \ac{MBH} is one year,
and the other parameters are set to $m=30 M_\odot$, $M=10^6 M_\odot$ $a=0.9$, $\alpha=(0.01,0.1)$, $p_0=12$ and $e_0=0.5$.
As shown in Fig. \ref{fig:waveform},
one can see that the two EMRI waveforms are indistinguishable at the initial stage. However, their distinction becomes prominent after six months of observation.
 Moreover, we can see that a larger coupling constant $\alpha$ leads to a more significant difference between the two kinds of  waveform.

To study the effects from various parameters on the
detection capability of LISA for
the EMRIs from the QCBHs, we plot the phase difference as a function of observation time by computing the Eq. \eqref{eq:dephsing}.
Following Refs. \cite{Datta:2019epe,Zi:2023omh}, the threshold value of dephasing is roughly taken as  $\delta\Psi^{\rm min}_{r,\phi}=1$ rad, above which the two kinds of signal can be resolved by LISA.
The azimuthal and radial dephasing as a function of observation time are plotted in Fig. \ref{fig:Dephasing:alpha},
where the deviation parameter of QCBH  with spin $a=0.9$ takes values as $\alpha=(0.0001,0.001,0.01,0.1,0.2,0.3)$.
As shown in Fig. \ref{fig:Dephasing:alpha}, the azimuthal and radial dephasing are growing when the deviation parameters $\alpha$ are increasing.
To evaluate the impact of orbital eccentricity on the dephasing,
we plot several examples of the initial orbital eccentricities $e_0=(0.1,0.3,0.5,0.7,0.9)$ in Fig. \ref{fig:Dephasing:ecc},
where the other parameters of QCBH are set as  $a=0.9$, $\alpha=0.001$ and $p_0=12$.
From the Figs. \ref{fig:Dephasing:alpha} and \ref{fig:Dephasing:ecc},
one can see that the deviation parameter has a more significant influence on the dephasing than the orbital eccentricity.
Note that the horizontal black dashed line represents the threshold value of dephasing in the Figs. \ref{fig:Dephasing:alpha} and \ref{fig:Dephasing:ecc}, so the region above this line means the EMRI waveforms with the quantum correction can be distinguished by LISA.

To quantitatively evaluate the effect of the trace anomaly on the EMRI waveform, the mismatch as a function of observation time is plotted in Fig. \ref{fig:mismatch} with $a=0.9$ and $p_0=12$. The left panel shows the mismatch with  various values of the eccentricity and a fixed $\alpha=0.01$, while the right panel shows the mismatch with various values of  deviation parameter and a  fixed eccentricity $e_0=0.5$.
In both panels the length of waveform is fixed as one year.
From the left panel of the figure, one can see that
it is possible to distinguish the quantum-corrected waveforms as the orbital eccentrics is bigger.
And the distinction between the two kinds of waveform is more evident when the initial orbital eccentricity is bigger.
From the right panel, the mismatch between waveforms emitted from the  different EMRI systems is sensitive to the deviation parameters,
and LISA has the potential to distinguish the quantum-corrected waveform with the deviation parameter
as small as $\alpha=0.003$.

\begin{figure*}[ht]
\centering
\includegraphics[width=0.69\textwidth]{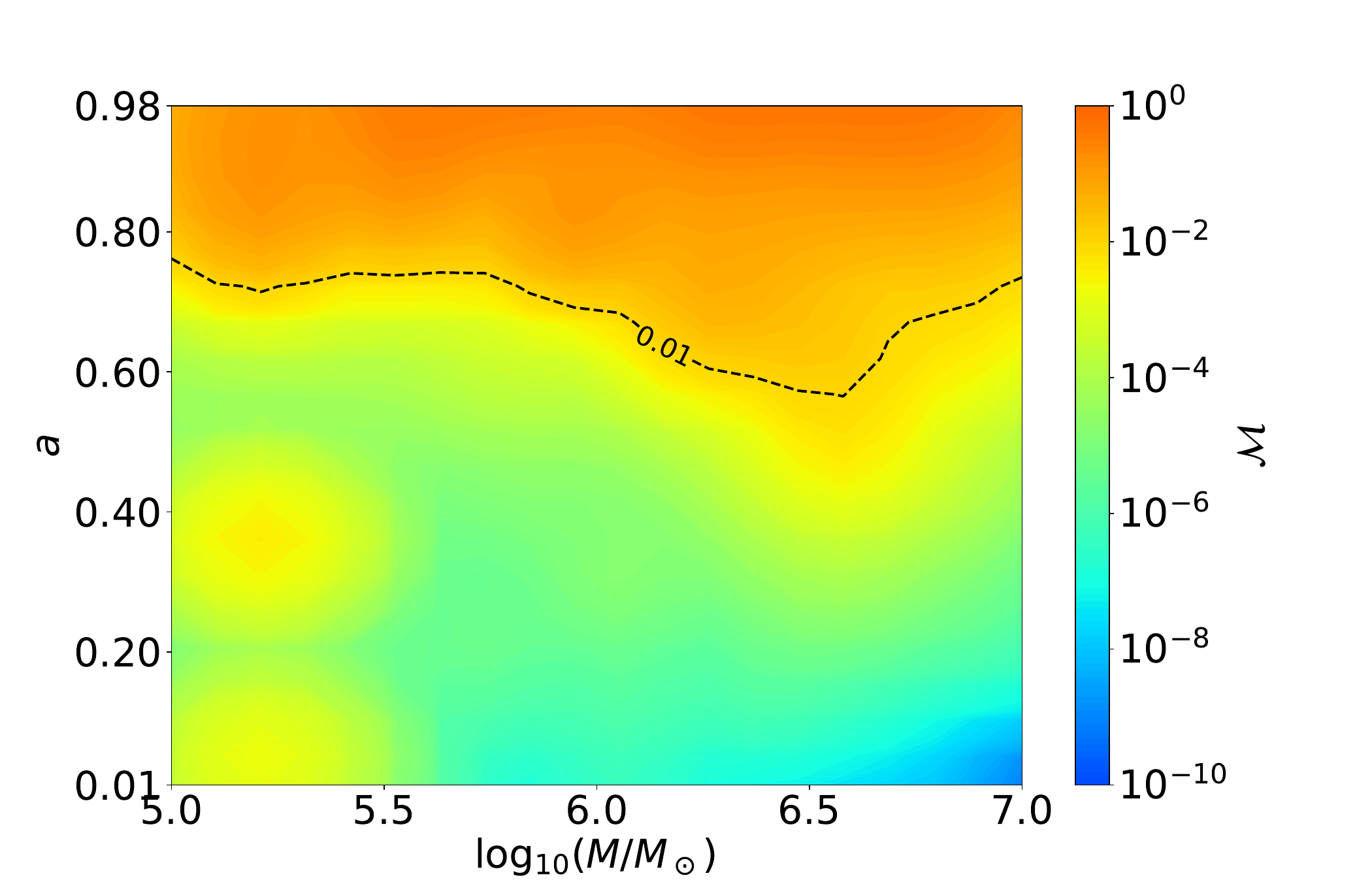}
\caption{Mismatch as a function of mass and spin of the QCBH with a deviation parameter $\alpha=0.01$ is plotted, where the horizontal black dashed line is the threshold value of mismatch $\mathcal{M}_{\rm min} =0.01$.
The other orbital parameters are set as $p_0=12$ and $e_0=0.5$.
}
\label{fig:mismatch:mass:spin}
\end{figure*}

Fig. \ref{fig:mismatch:mass:spin} shows the mismatch as a function of mass and spin of the MBH with the deviation parameter $\alpha=0.01$ and the initial orbital parameters $p_0=12$ and $e_0=0.5$, considering the one year observation of the EMRI signal.
The horizontal black dashed line is the threshold value of the mismatch $\mathcal{M}=0.01$.
From this figure, we can find that the mismatch is more sensitive to the spin of the MBH than to its mass.  When the spin of the MBH satisfies $a\gtrsim0.79$, the EMRI waveform with the presence of the  trace anomaly can be distinguished by LISA.

\subsection{Constraint on the trace anomaly}\label{Constraint}
In this subsection, we perform the parameter estimation
for the deviation parameter $\alpha$ using the FIM method. By setting the central values of the  $\alpha$ to a given value, we can obtain the constraint on the strength of trace anomaly by LISA.
 In general, the inspiral is truncated  at the
last stable orbit of the MBH \cite{Stein:2019buj}. In this work, by  choosing the initial values of the orbital parameters appropriately, we can fix the length of the waveform to one year and keep the inspiral way from the cutoff. The initial orbital parameters are set as $p_0=12$ and $e_0=0.5$, and the deviation parameter of the MBH is fixed as $\alpha=0.001$.
Here we only focus on the effects from the  mass and spin of the MBH and let other parameters fixed, since the former ones are more significant.

\begin{figure*}[ht]
\centering
\includegraphics[width=0.69\textwidth]{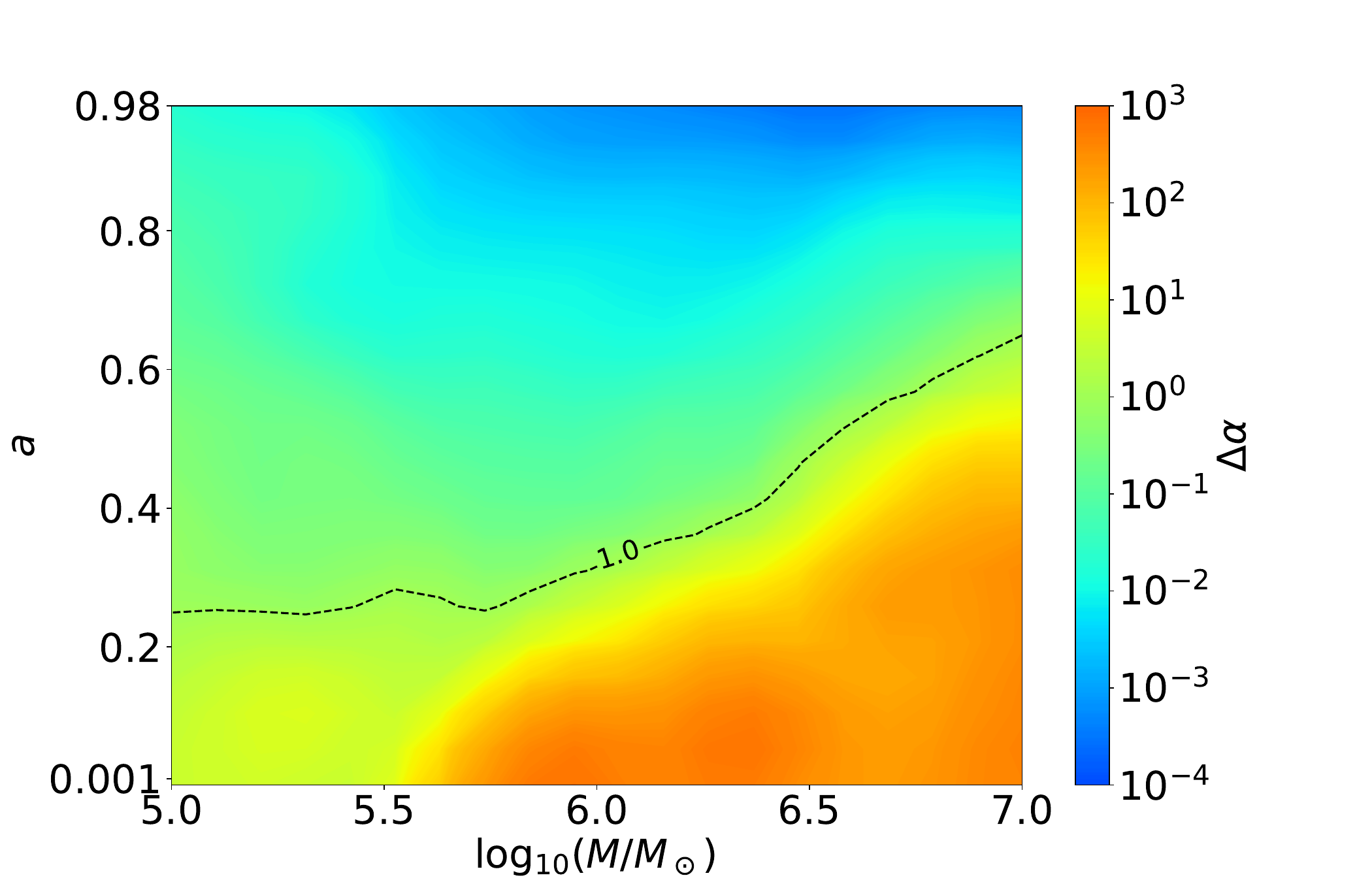}
\caption{Constraint on the deviation parameter $\Delta \alpha$ as a function of mass and spin of the MBH with  $\alpha=0.01$ is plotted, where the black dashed line denotes the threshold value of the measurement error $\Delta\alpha_{\rm max} =1.0$. The initial orbital parameters are set as $p_0=12$ and $e_0=0.5$.}
\label{fig:fim}
\end{figure*}

We plot the parameter estimation uncertainty $\Delta\alpha$ as a function of spin and  mass of the MBH in Fig. \ref{fig:fim}.
It is found that the uncertainty of
the deviation parameter $\Delta \alpha$ decreases with the spin parameter when the mass of the MBH is fixed. In particular, if the spin of the MBH is small, the uncertainty could exceed the threshold 1, which means $\alpha$ cannot be constrained with EMRI signals in this case.
However, the parameter estimation accuracy gets improved quickly with the increase of the spin of the MBH. So the MBH
with the largest spin has the best constraint on $\alpha$. We can find that when the spin parameter $a>0.8$ and the mass of the MBH take values around $10^{6}M_\odot$, the constraint on the deviation parameter can reach the level of $\sim10^{-4}$ for one year observation by LISA.

\begin{figure*}[ht]
\centering
\includegraphics[width=0.56\textwidth]{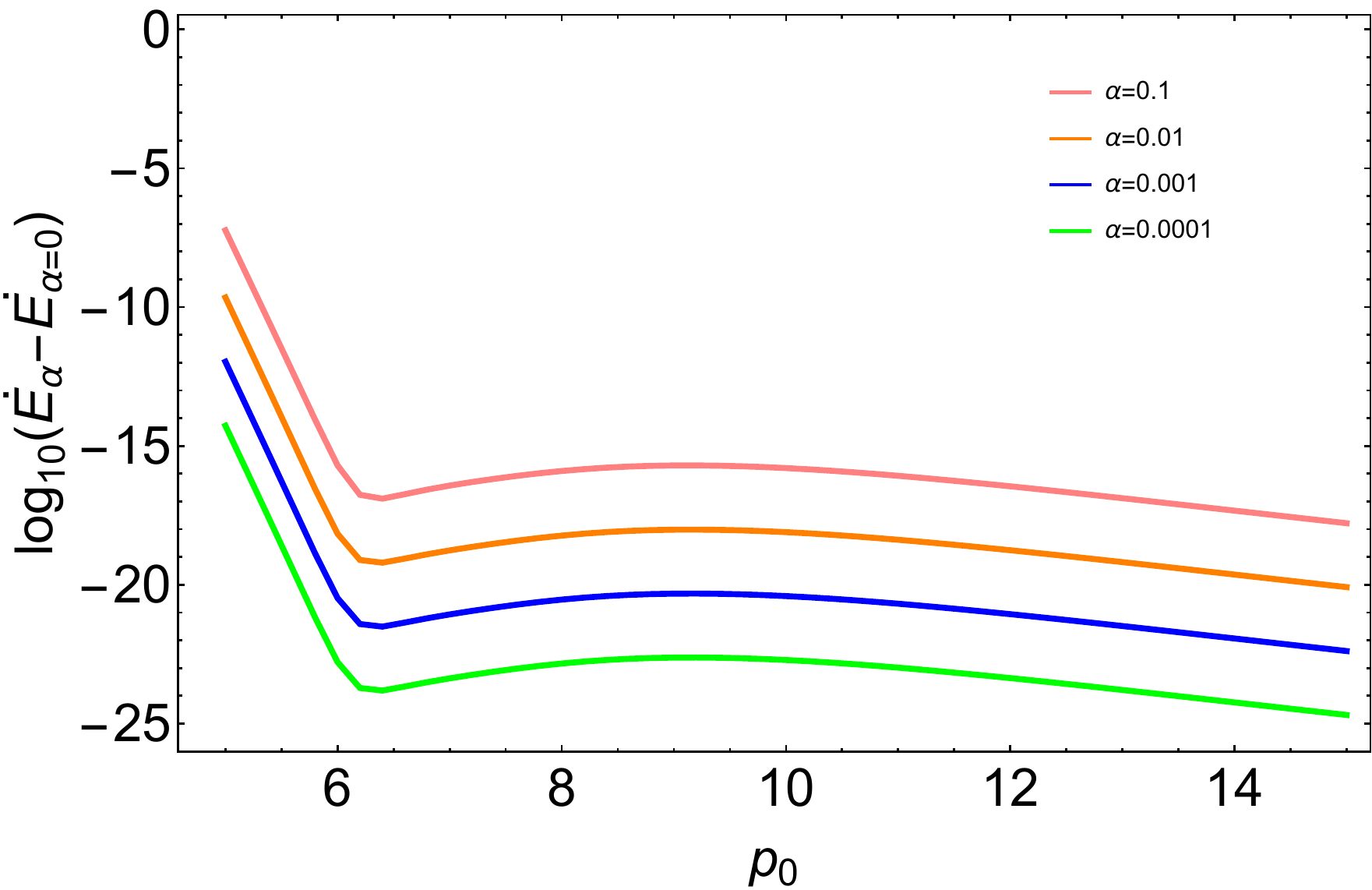}
\caption{The logarithm of difference values of energy fluxes for the QCBH and Kerr BH as a function of the initial orbital semi-latus rectum  for several cases of quantum-corrected parameters $\alpha=(0.0001,0.001,0.01,0.1)$ are plotted. The other orbital parameters are set as $a=0.9$ and $e_0=0.5$.}
\label{fig:flux:difference}
\end{figure*}

\section{Conclusion}\label{summary}
In this paper, we computed the gravitational waves emitted from the extreme mass ratio inspirals around rotating BHs in semiclassical gravity sourced by the trace anomaly  using the kludge approximate method.
We firstly derived the orbital energy, angular momentum and fundamental orbital frequencies for orbits on the equatorial plane. We found that, for the gravitational radiation described by quadrupole formulas, the contribution from the trace anomaly only appears at  higher order terms in the the energy flux comparing with the standard Kerr case. Therefore, we evaluated the hybrid orbital evolution equations using the energy flux of Kerr BH, then computed the EMRI waveform with the quadruple formula \cite{Barack:2003fp,Barsanti:2022ana}. Moreover,
we assessed the differences of phase of the waveforms with and without the trace anomaly by computing the dephasing.
Our results indicated that the dephasing is more significant for the larger values of the initial eccentricity $e_0$ and the deviation parameter $\alpha$.
By computing mismatches of the two kinds of waveform,
we found that QCBH with  $a\gtrsim 0.79$ can be distinguished from the Kerr BH.
According to the parameter estimation of quantum-corrected  parameter $\alpha$,
the constraints on the parameter $\alpha$ are seriously subjected with the spins of QCBH.
In particular, LISA can determine the parameter $\alpha$ within a fractional error of $\sim10^{-4}$ for the higher spinning QCBH, and EMRI sources with the lower spinning QCBH could not be measurable, especially the more massive QCBH.

It is interesting to perform the study using the complete perturbation  theory for the QCBHs, which can provide more accurate EMRI waveforms for the evaluation of the detection ability of the trace anomaly.  However, since the horizon of the QCBH is non-spherical, one can expect that the scheme would be much difficult than the Kerr case.

\appendix
\section{Energy flux}\label{appA:flux}
In this section we introduce the quadrupole formulas of energy flux  derived by Peters
and Mathews \cite{Peters:1963ux,Peters:1964zz}, where the energy flux  can be written as
\begin{equation}
\dot{E} =\frac{1}{5m}\left( \frac{d^3I_{ij}}{dt^3} \frac{d^3I^{ij}}{dt^3}
-\frac{d^3I_i^i}{dt^3} \frac{d^3I^j_j}{dt^3},
\right)
\end{equation}
with the quadrupole moment tensor of mass $I_{ij}$, which can be given by
\begin{equation}
I_{ij} = m x_ix_j,
\end{equation}
where $x_i$ is  the  position vector between the smaller object and MBH.
Under the weak field approximate condition, we can obtain the formulas of quadrupole moment tensor $Q_{ij}$
and energy flux $\dot{E}$ via the intricate algebraical computation.
To illustrate the modification effect of the deviation parameter $\alpha$ on the EMRI energy flux,
we plot the logarithmic differences of the energy fluxes from the QCBH and Kerr BH in Fig. \ref{fig:flux:difference},
the symbols $\dot{E}_\alpha$ and  $\dot{E}_{\alpha=0}$ denote to the energy flux of QCBH and Kerr BH.
From the Fig. \ref{fig:flux:difference}, one can see that the correction from QCBH on EMRI fluxes can be
ignored comparing to the case of the Kerr BH.
This is because that the contribution of the deviation parameter $\alpha$  appears at  higher order of the weak-field expansion of Eq. \eqref{ADM:mass}, that is
\begin{equation}
\mathcal{M}(r,\theta) = M+\frac{2M^2\alpha}{r^3}+\mathcal{ O}\left(\frac{1}{r^5}\right).
\end{equation}
Therefore, the evolution of orbital parameter in the QCBH spacetime can be approximately addressed
with the fluxes of Kerr BH, we adopt the analytic fluxes of energy and angular moment developed by
Gair and  Glampedakis in Ref. \cite{Gair:2005ih}.
The fluxes can be read with our symbol as follows
\begin{eqnarray}
\dot{E}&=& -\frac{32}{5} \frac{\mu^2}{M^2} \left(\frac{M}{p}\right)^5
(1-e^2)^{3/2}\left [ g_1(e) -q\left(\frac{M}{p}\right)^{3/2} g_2(e)
\nonumber
\right. \\&& \left.  -\left(\frac{M}{p}\right) g_3(e) +
\pi\left(\frac{M}{p}\right)^{3/2} g_4(e)
\right. \\&& \left. -
\left(\frac{M}{p}\right)^2 g_5(e)+ q^2\left(\frac{M}{p}\right)^2 g_6(e) -
\nonumber
\right. \\&& \left.
\pi\left(\frac{M}{p}\right)^{5/2} g_7(e)
+ q\left(\frac{M}{p}\right)^{5/2} g_8(e) \right ] ,\
\label{new_Edot}
\nonumber \\
\dot{L}_z&=& -\frac{32}{5} \frac{\mu^2}{M} \left(\frac{M}{p}\right)^{7/2}
(1-e^2)^{3/2}
\left [g_9(e)
\right.\nonumber \\
&& \left. -q\left(\frac{M}{p}\right)^{3/2} g_{10}(e) -\left(\frac{M}{p}\right) g_{11}(e)
\right.
\nonumber \\
&& \left.
+\pi\left(\frac{M}{p}\right)^{3/2} g_{12}(e)
-\left(\frac{M}{p}\right)^2 g_{13}(e)
\right.\nonumber \\&& \left.
+ q^2\left(\frac{M}{p}\right)^2 g_{14}(e)
-\pi\left(\frac{M}{p}\right)^{5/2} g_{15}(e)
\right.\nonumber \\&& \left.
+ q\left(\frac{M}{p}\right)^{5/2} g_{16}(e) \right ] ,\
\label{new_Ldot}
\end{eqnarray}
where the $e$-dependent coefficients are,
\begin{eqnarray}
g_1(e) &=& 1 + \frac{73}{24} e^2  + \frac{37}{96}e^4, \nonumber\\
g_2(e) &=& \frac{73}{12} + \frac{823}{24} e^2
 + \frac{949}{32}e^4
+ \frac{491}{192}e^6 ,\nonumber \\
g_3(e) &=& \frac{1247}{336} + \frac{9181}{672} e^2 , \nonumber\\
g_4(e) & =& 4 + \frac{1375}{48} e^2  ,\nonumber\\
g_5(e) &=& \frac{44711}{9072} + \frac{172157}{2592} e^2 ,\nonumber\\
g_6(e) &=& \frac{33}{16} + \frac{359}{32} e^2 ,\nonumber\\
g_7(e) &=& \frac{8191}{672} + \frac{44531}{336} e^2  ,\nonumber\\
g_8(e) &=& \frac{3749}{336} - \frac{5143}{168} e^2    ,\nonumber\\
g_9(e) &=& 1 + \frac{7}{8} e^2  , \nonumber\\
g_{10}(e) & = &\frac{61}{12} + \frac{119}{8} e^2 +\frac{183}{32}e^4 ,\nonumber\\
g_{11}(e) &=& \frac{1247}{336} + \frac{425}{336} e^2, \nonumber \\
g_{12}(e) &= & 4 + \frac{97}{8} e^2 ,\nonumber\\
g_{13}(e) &=& \frac{44711}{9072} + \frac{302893}{6048} e^2 ,\nonumber\\
g_{14}(e) &=& \frac{33}{16} + \frac{95}{16} e^2 ,\nonumber\\
g_{15}(e) &=& \frac{8191}{672} + \frac{48361}{1344} e^2 ,\nonumber\\
g_{16}(e)& =& \frac{417}{56} - \frac{37241}{672} e^2 .\
\end{eqnarray}

\section{Stability of the Fisher matrix}\label{appB:Stability}
In this appendix, following the method in Ref. \cite{Piovano:2021iwv,Zi:2023pvl},
we compute the stability of the covariance matrix with the EMRI waveforms.
This can be achieved as follows:
firstly compute small perturbations of components of Fisher matrices, then see the behavior of covariance matrices. The stability can be characterized quantitatively by the following equation
\begin{equation}\label{fim:stab}
	\delta_{\rm stability} \equiv \mathbf{max}_{\rm ij} \left[\frac{((\Gamma+F)^{-1} - \Gamma^{-1})^{ij}}{(\Gamma^{-1})^{ij}}\right]
\end{equation}
where $F_{ij}$ is the deviation matrix, the elements is a uniform distribution $U\in[a, b]$.
To assess the stability of FIM with the EMRI signal modified by QCBH, we list the results
of  stability $\delta_{\rm stability}$in the Table \ref{FIMstability}.

\begin{table*}[!htbp]
	\caption{The stability $\delta_{\rm stability}$ of FIM with EMRI waveforms from the QCBH with mass $M=10^{6}M_\odot$ and $\alpha=0.01$ is computed, including the different spinning caes.	}\label{FIMstability}
	\begin{center}
		\setlength{\tabcolsep}{5mm}
		\begin{tabular}{|c|c|c|c|c|c|}
			\hline
			\multirow{2}{*}{$U$}& \multicolumn{5}{|c|}
			{spin $a$}\\
			\cline{2-6}
			& $0.1$ &$0.3$ &$0.5$ &$0.7$ & $0.9$ \\
            \hline
			$\in[-10^{-6},10^{-6}]$ & $3.47\times10^{-1}~~$ &$2.11\times10^{-1}$ &$6.82\times10^{-2}$
			&$3.65\times10^{-2}$  &$2.72\times10^{-4}$
              \\		
			\hline
			$\in[-10^{-9},10^{-9}]$ & $9.64\times10^{-2}~~$ &$3.82\times10^{-3}$ &$7.16\times10^{-4}$
			&$3.07\times10^{-4}$  &$1.35\times10^{-4}$
             \\
			\hline	

		\end{tabular}
	\end{center}
\end{table*}

\section*{Acknowledgments}
The work is in part supported by NSFC Grant
No.12205104, ``the Fundamental Research Funds for the Central Universities'' with Grant No.  2023ZYGXZR079, the Guangzhou Science and Technology Project with Grant No. 2023A04J0651 and the startup funding of South China University of
Technology. T. Z. is also funded by China Postdoctoral Science Foundation Grant
No. 2023M731137.


%

\end{document}